\documentstyle[12pt]{article}
\def\be {\begin{equation}}
\def\ee {\end{equation}}
\def\ba {\begin{eqnarray}}
\def\ea {\end{eqnarray}}

\def\bi {\begin{itemize}}
\def\ei {\end{itemize}}
\begin{document}
\def\bea{\begin{eqnarray}}
\def\eea{\end{eqnarray}}
\title{\bf {Non-Rotating BTZ Black Hole Area Spectrum from Quasi-normal Modes }}
 \author{M.R. Setare  \footnote{E-mail: rezakord@ipm.ir}
  \\{Physics Dept. Inst. for Studies in Theo. Physics and
Mathematics(IPM)}\\
{P. O. Box 19395-5531, Tehran, IRAN }}
\date{\small{}}

\maketitle
\begin{abstract}
In this paper, by using the quasi-normal frequencies for
non-rotating BTZ black hole derived by Cardos and Lemos, also via
Bohr-Sommerfeld quantization for an adiabatic invariant, $I=\int
{dE\over \omega(E)}$, which $E$ is the energy of system and
$\omega(E)$ is vibrational frequency, we leads to an equally
spaced mass spectrum. The result for the  area of event horizon
is $A_{n}=2\pi \sqrt{\frac{nm \hbar}{\Lambda}}$ which is not
equally spaced, in contrast with area spectrum of black hole in
higher dimension.
 \end{abstract}
\newpage

 \section{Introduction}
 Any non-dissipative systems has modes of vibrations, which forming a
complete set, and called normal modes. Each mode having a given
real frequency of oscillation and being independent of any other.
The system once disturbed continues to vibrate in one or several
of the normal modes. On the other hand, when one deals with open
dissipative system, as a black hole, instead of normal modes, one
considers quasi-normal modes for which the frequencies are no
longer pure real, showing that the system is loosing energy. In
asymptotically flat spacetimes the idea of QNMs started with the
work of Regge and Wheeler \cite{reggeW} where the stability of a
black hole was tested, and were first numerically computed by
Chandrasekhar and Detweiler several years later \cite{Chandra1}.
The quasi-normal modes bring now a lot of interest in different
contexts: in AdS/CFT correspondence
\cite{Horowitz-Habeny}-\cite{Moss-Norman}(according to the
AdS/CFT conjecture \cite{mal}, the black hole corresponds to a
thermal state in the conformal field theory, and the decay of the
test field in the black hole spacetime corresponds to the decay
of the perturbed state in the CFT.  The dynamical timescale for
the return to thermal equilibrium is very hard to compute
directly, but can be done relatively easily using the AdS/CFT
correspondence), when considering thermodynamic properties of
black holes in loop quantum gravity \cite{LQG}-\cite{mot}, in the
context of possible connection with
critical collapse \cite{Horowitz-Habeny,BHCC,kim}.\\
The quantization of the black hole horizon area is one of the most
interesting manifestations
 of quantum gravity. Since its first prediction by
Bekenstein  in 1974 \cite{bek1}, there has been much work on this
topic \cite{bek2}-\cite{bir}. A characteristic feature of loop
quantum gravity is the  discrete spectrum of several geometrical
quantities \cite{1,2}. In the three spacetime dimensions, the
length operator plays a role analogous to the area operator in
four dimensional spacetime. For the 3D Euclidean spacetime, the
eigenvalues of the length are discrete \cite{3}, in the other hand
the spectrum of the length in the 3D Lorentzian case is continuous
\cite{4}.
\\  Recently, the quantization of the
black hole area has been considered \cite{hod, LQG} as a result of
the absorption of a quasi-normal mode excitation. Bekenstein's
idea for quantizing a black hole is based on the fact that its
horizon area, in the nonextreme case, behaves as a classical
adiabatic invariant \cite{bek1}, \cite{bek3}. In the spirit of
Ehrenfest principle, any classical adiabatic invariant corresponds
to a quantum entity with discrete spectrum, Bekenstein conjectured
that the horizon area of a non extremal quantum black hole should
have a discrete eigenvalue spectrum. Moreover, the possibility of
a connection between the quasinormal frequencies of black holes
and the quantum properties of the entropy spectrum was first
observed by Bekenstein \cite{bek4}, and further developed by Hod
\cite{hod}. In particular, Hod proposed that the real part of the
quasinormal frequencies, in the infinite damping limit, might be
related via the correspondence principle to the fundamental
quanta of mass and angular momentum.\\
In this paper we would like to obtain the area spectrum of
non-rotating BTZ black hole. Using the quasi-normal frequencies
which has been derived by Cardos and Lemos in
\cite{CFT-Correspond}, also via Boher-Sommerfeld quantization for
an adiabatic invariant as $I=\int {dE\over \omega(E)}$, which $E$
is the energy of system and $\omega(E)$ is vibrational frequency,
we leads to an equally spaced mass spectrum. We show that the
results for the spacing of the area spectrum differ from black
holes in $3+1$ dimension, in fact we have not found a
quantization of horizon area.

\section{Non-rotated BTZ black hole and quasi-normal modes}
The black hole solutions of Ba\~nados, Teitelboim and Zanelli
\cite{banados1,banados2} in $(2+1)$ spacetime dimensions are
derived from a three dimensional theory of gravity \be S=\int
dx^{3} \sqrt{-g}\,({}^{{\small(3)}} R+2\Lambda) \ee with a
negative cosmological constant ($\Lambda=\frac{1}{l^2}>0$).
\par\noindent
The corresponding line element is \be ds^2 =-\left(-M+
\frac{r^2}{l^2} +\frac{J^2}{4 r^2} \right)dt^2
+\frac{dr^2}{\left(-M+ \displaystyle{\frac{r^2}{l^2} +\frac{J^2}{4
r^2}} \right)} +r^2\left(d\theta -\frac{J}{2r^2}dt\right)^2
\label{metric}\ee with $M$ the Arnowitt-Deser-Misner (ADM) mass,
$J$ the angular momentum (spin)
 of the BTZ black hole and $-\infty<t<+\infty$, $0\leq r<+\infty$, $0\leq \theta <2\pi$.
\par \noindent
In this paper we would like to consider the quasi-normal modes of
the non-rotated (uncharged) BTZ black holes. The non rotated BTZ
black hole can be obtain by putting $J=0$ in Eq.(\ref{metric}) as
following \be ds^2 =-(-M+ \frac{r^2}{l^2}  )dt^2 +\frac{dr^2}{(-M+
\frac{r^2}{l^2})}+r^{2} d\theta ^2, \label{metric2}\ee which has
an horizon at \be r_{+}=\sqrt{\frac{M}{\Lambda}}, \label{hori} \ee
and is similar to Schwarzschild black hole with the important
difference that it is not asymptotically flat but it has constant
negative curvature. The temperature of the event horizon is given
by \be T_{H}=\frac{\sqrt{\Lambda M }}{2\pi}. \label{tem} \ee The
entropy of non-rotating BTZ black hole is as following \be
S_{bh}=4\pi r_{+}.\label{entro} \ee For BTZ black hole with $J=0$,
the free energy is given by \cite{myang} \be F=-M \label{free},
\ee also the relation between energy and free energy is as
following \be E=F+T_{H}S_{bh} \label{eneq}. \ee From
Eqs.(\ref{hori}-\ref{eneq}) one can obtain \be E=M \label{masseq}
\ee The quasi-normal frequencies for non-rotating BTZ black hole
have been obtained by Cardoso and Lemos in \cite{CFT-Correspond}
\be \omega=\pm m-2iM^{1/2}(n+1) \hspace{1cm}
n=0,1,2,...\label{quas} \ee where $m$ is the angular quantum
number. Let $\omega=\omega_{R}-i\omega_{I}$, then
$\tau=\omega_{I}^{-1}$ is the effective relaxation time for the
black hole to return to a quiescent state. Hence, the relaxation
time $\tau$ is arbitrary small as $n\rightarrow \infty$. We assume
that this classical frequency plays an important role in the
dynamics of the black hole and is relevant to its quantum
properties \cite{{hod},{LQG}}. In particular, we consider
$\omega_{R}$ the real part of $\omega$, to be a fundamental
vibrational frequency for a black hole of energy $E=M$. Given a
system with energy $E$ and vibrational frequency $\omega$ one can
show that the quantity \be I=\int {dE\over \omega(E)},
\label{bohr} \ee is an adiabatic invariant \cite{kun}, which via
Bohr-Sommerfeld quantization has an equally spaced spectrum in the
semi-classical (large $n$) limit: \be I \approx n\hbar.
\label{smi} \ee Now by taking $\omega_{R}$ in this context, we
have \be I=\int \frac{dE}{\omega_{R}}=\int \frac{1 }{\pm m}dM =
\frac{M}{\pm m}+c, \label{adin} \ee where $c$ is a constant,
therefore the mass spectrum is equally spaced
\begin{eqnarray}
&&
M=mn\hbar, \hspace{0.5cm}m\geq 0 \nonumber \\
&&  M= -mn\hbar , \hspace{0.5 cm}m<0 \label{massspec}.
\end{eqnarray} Then for mass spacing we have \be \Delta M=\pm
m\hbar, \label{spacing} \ee which is the fundamental quanta of
black hole mass. As one can see the mass spectrum is equally
spaced only for a fixed $m$. For different $m$ there are
multiplets with different values of spacing which is given by
Eq.(\ref{spacing}). On the other hand, the black hole horizon area
is given by \be A=2\pi r_{+}. \label{area1} \ee Using
Eq.(\ref{hori}) we obtain \be A=2\pi \sqrt{\frac{M}{\Lambda}}.
\label{area2} \ee The Bohr-Sommerfeld quantization law and
Eq.(\ref{adin}) then implies that the area spectrum is as
following, \be A_{n}=2\pi \sqrt{\frac{nm \hbar}{\Lambda}}
.\label{spec} \ee As one can see although the mass spectrum is
equally spaced but the area spectrum is not equally spaced.
  \section{Conclusion}
In this paper we have considered the non-rotating BTZ black hole,
with the knowledge of the quasi-normal mode spectrum as given in
\cite{CFT-Correspond}, we have prescribed how their horizon area
would be quantized?  In \cite{bir3} Birmingham et al have shown
that the quantum mechanics of the rotating BTZ black hole is
characterized by a Virasoro algebra at infinity. Identifying the
real part of the quasi-normal frequencies with the fundamental
quanta of black hole mass and angular momentum, they found that
an elementary excitation corresponds exactly to a correctly
quantized shift of the Virasoro generator $L_{0}$ or $\bar{L_0}$
in this
algebra.\\
Via Boher-Sommerfeld quantization for an adiabatic invariant as
$I=\int {dE\over \omega(E)}$, which $E$ is the energy of system
and $\omega(E)$ is vibrational frequency, we leads to an equally
spaced mass spectrum. Similar to \cite{bir3}we have not found a
quantization of horizon area. The result for the  area of event
horizon is $A_{n}=2\pi \sqrt{\frac{nm \hbar}{\Lambda}}$ which is
not equally spaced, in contrast with area spectrum of black hole
in higher dimension. This result is a bit difficult to interpret
physically since the quasi-norma mode spectrum in $2+1$ is quite
different from what it is in higher dimensions
  \vspace{3mm}
\section*{Acknowledgement }
I would like to thank Prof. Gabor Kunstatter for reading the
manuscript.

\end{document}